\documentstyle[pra,aps,eqsecnum]{revtex}
\preprint{NSF-ITP-97-017}
\draft
\title{Time Dependent Floquet Theory\\ and
Absence of an Adiabatic Limit}
\author{
Daniel W.~Hone$^{1,2}$~\footnote{Electronic address:
   hone@itp.ucsb.edu},
Roland Ketzmerick$^{2,3}$~\footnote{Electronic address:
    roland@chaos.gwdg.de},
and
Walter Kohn$^{1}$~\footnote{Electronic address:
   kohn@physics.ucsb.edu}
}
\address{
$^{1}$ Physics Department, UCSB, Santa Barbara, CA 93106 \\
$^{2}$ Institute for Theoretical Physics, UCSB, Santa Barbara, CA 93106\\
$^{3}$ MPI f\"ur Str\"omungsforschung
und Institut f\"ur Nichtlineare Dynamik der Universit\"at G\"ottingen, 
Bunsenstr.~10, D-37073 G\"ottingen, Germany}
\date{\today}
\narrowtext

\begin{document}

\maketitle

\begin{abstract}
Quantum systems subject to time periodic fields of  {\em finite} 
amplitude $\lambda$ have conventionally been handled either by
low order perturbation theory, for $\lambda$ not too large, or
by exact diagonalization within a finite basis of $N$ states.
An adiabatic limit, as $\lambda$ is switched on arbitrarily
slowly, has been assumed.  But the validity of these procedures 
seems questionable in view of the fact that, as $N \rightarrow
\infty$, the quasienergy spectrum becomes dense, and numerical
calculations show an increasing number of weakly avoided 
crossings (related in perturbation theory to high order resonances).
This paper deals with the highly non-trivial
behavior of the solutions in this limit. 
The Floquet states, and the associated quasienergies, become highly
irregular functions of the amplitude $\lambda$. The
mathematical radii of convergence of perturbation theory in $\lambda$ 
approach zero. There is no adiabatic limit of the wave functions
when $\lambda$ is turned on arbitrarily slowly.  However, the
quasienergy becomes independent of time in this limit.  We introduce 
a modification of the adiabatic theorem.  We
explain why, in spite of the pervasive pathologies of the
Floquet states in the limit $N \rightarrow \infty$, the
conventional approaches are appropriate in almost all
physically interesting situations.

\end{abstract}
\pacs{42.50.Hz, 42.65.Vh, 03.65.-w, 05.45.+b}


\section{ Introduction}

Physical systems subject to finite time periodic perturbations
of amplitude $\lambda$ and period $T\equiv 2\pi/\omega$ have been
extensively studied~\cite{chu} by making use of the Floquet 
theorem.  This theorem, a consequence of the discrete time 
translation symmetry
of the Hamiltonian, states that there is a complete set of 
quasiperiodic solutions of the time dependent Schr\"odinger equation 
which, when $t\rightarrow t+T$, are simply multiplied by a phase
factor $\exp (-i\epsilon_n T)$, where $\epsilon_n$ is called the
``quasienergy".  This phase factor defines $\epsilon_n$ only
modulo $\omega$, and so the quasienergy may always be taken to lie 
in the strip $0\le\epsilon_n <\omega$.  

For such systems the quasienergies $\epsilon_n$ are of
comparable interest to the energy levels of
time independent systems. One would expect
to use the subscripts $n$ as unambiguous labels of the 
time dependent states
of a periodically driven system as the magnitude $\lambda$
of the perturbation is switched on
adiabatically, as in the case of time
independent quantum systems.  Indeed, there is a substantial 
literature studying
Floquet systems along these lines, usually employing
numerical methods, in a {\em finite} set of
basis states \cite{manypapers}.  Of special interest are the 
``avoided crossings" (ac's) --- regions in the $\lambda - 
\epsilon$ plane where
two quasienergies approach each other as a function of $\lambda$
and (except for special symmetries) avoid
crossing one another.

But there is a difference in principle when a complete {\em infinite} 
set of basis states is included.   For a spatially
confined system, with an infinite number of discrete energy
levels for $\lambda = 0$, there is also
an infinite number of quasienergies for $\lambda > 0$,
and the spectrum  fills the fundamental strip densely. 
In fact, typically it is a dense point spectrum \cite{howland}.
(For example, in the special case of a particle in a one 
dimensional square
well and vanishing $\lambda$, Weyl\cite{weyl}
has shown that the energy eigenvalues fill the fundamental
strip densely and uniformly).   As the number 
of basis states becomes infinite there is a weakly avoided
crossing (hereafter ``wac") near each point in the 
$\lambda - \epsilon$ strip.
This leads to qualitatively new issues:  Do the Floquet
state solutions for a given $\lambda$ converge to 
well defined limits?  Are the Floquet states and quasienergies 
well behaved functions of $\lambda$?  And 
is there a well defined limiting path which the system follows as
$\lambda$ is switched on arbitrarily slowly --- {\it i.e.},
does an adiabatic limit exist?   We 
have examined these questions in a variety of approximate ways
and have arrived at a coherent picture, though generally 
we do not have conclusive mathematical proofs.  
We find that the radii of convergence of power
series expansions in $\lambda$, starting from the unperturbed
eigenstates, are zero. 
Floquet states and their quasienergies 
are discontinuous functions of $\lambda$ everywhere; there is 
no adiabatic limit
in the usual sense (though we will propose a useful weakened
modification of the adiabatic theorem). At the same time, in the
limit of slow switching on of $\lambda$ the quasienergy 
remains arbitrarily close to 
its initial ($\lambda = 0$) value.  
We will explain the consistency of these results with the
well established success of standard time dependent perturbation 
theory and of the adiabatic theorem. 

We note that these features will not be seen directly in any
numerical study as $N$ becomes large.  The effects become
increasingly weak very rapidly as the basis size is increased.
Although, for a given interval in $\lambda$, inclusion of very 
high lying levels does have a major impact on Floquet states and
quasienergies, it is only over an increasingly smaller range of
$\lambda$, and this becomes at some stage unobservable on the 
scale of numerical accuracy available to the computer.

In Section II we review elementary Floquet theory in a 
{\em finite} basis.  The problems arising for an infinite basis
are discussed in Section III.  In Section IV we analyze the convergence of
the Floquet states as the size of the basis becomes infinite.  
Section V is devoted to the
dependence of the states and quasienergies on $\lambda$, 
including questions of labelling
Floquet states, and of the existence of an adiabatic limit, 
when $N\rightarrow\infty$.  The relationship to finite
order time dependent perturbation theory is discussed in 
Section VI.  Our conclusions are summarized in Section VII.

\section{Floquet Theory in a Finite Basis}

In this section we review the elements
of time periodic Hamiltonian systems,
\begin{equation}
H(t) = H_0 + \lambda H_1(t) \;\; {\rm with} \;\; H_1(t+T)=H_1(t) ,
\label{Hperiodic}
\end{equation}
when approximated by a {\em finite} matrix, $H^N(t)$, in the representation
of the $N$ lowest states of the time independent Hamiltonian $H_0$.
In the next section we will discuss the considerable problems
that occur in the limit $N \rightarrow \infty$.

Because of the discrete time translational symmetry of $H^N(t)$,
there is a complete set of solutions $\varphi^N(t)$ of the 
time dependent Schr\"odinger equation (with $\hbar = 1$),
\begin{equation}
i \frac{d\varphi^N}{dt} = H^N(t) \varphi^N,
\label{SE}
\end{equation}
which are of the Floquet form,
\begin{equation}
\varphi^{N}_n(t) = \exp(-i \epsilon_n^{N} t)
u_n^{N}(t),
\label{Floquet}
\end{equation}
with a time periodic part 
\begin{equation}
u_n^{N}(t+T) = u_n^{N}(t); \qquad (n=1,2,\dots,N) ,
\label{perpart}
\end{equation}
and ``quasienergies" $\epsilon_n^{N}$ which may be taken to lie
in the interval $[0,\omega)$,
with $\omega=2\pi/T$.
For given $\lambda$ one may label the Floquet states
in order of increasing quasienergy, the
eigenvalue which characterizes discrete time translational symmetry.

Figure 1 shows such a quasienergy spectrum (for finite $N$)
as a function of $\lambda$
for the example of a free particle in a one-dimensional
box ($x \in [-a,a]$) with harmonic driving, 
\begin{equation}
H(t) = p^2/2m + \lambda \sin(\frac{\pi x}{2a}) \cos(\omega t) .
\end{equation}
The quasienergies are continuous functions of $\lambda$ that do not cross,
but show avoided crossings (ac's) provided there are no symmetries
which allow actual crossings. These ac's are abundant in 
the spectra of time periodic systems
and are of central interest in this work.  As the basis size is
increased, the newly introduced avoided crossings tend to become
rapidly weaker (smaller gaps at the crossing and smaller range
of $\lambda$ over which their effects are substantial).  They
therefore tend to become isolated from one another, and it is
useful and meaningful to consider their effects individually,
as we do in the following.

As two quasienergy lines pass an isolated weakly avoided crossing
(wac) at $\lambda=\lambda_1$,
the corresponding Floquet states rapidly interchange their forms.
At the point $\lambda=\lambda_1$ they  
are very nearly
linear combinations of the two Floquet functions just outside the
region of the ac, with amplitudes of equal magnitude.
As $|\lambda -\lambda_1|$ grows, this mixing decreases rapidly,
as follows.
The functions are mixed
by no more than a given small relative weight
$\delta$, provided that $|\lambda - \lambda_1|>w(\delta)/2$, where
\begin{equation}
w(\delta)=\frac{\Delta\epsilon}{\delta |\sigma_1-\sigma_2|} .
\label{width}
\end{equation}
Here $\Delta\epsilon$ is the quasienergy splitting at the 
ac, and $\sigma_1$ and $\sigma_2$ are the  slopes
$ d\epsilon / d\lambda $ of the quasienergies 
at the crossing point in the absence of the
terms connecting these states (see Fig.~1).
We neglect the weak influence of all other states near this wac.

Using standard Floquet state perturbation theory\cite{sambe} in 
$\lambda$, one finds that 
the second order term describes very well the overall behaviour of
the quasienergies for small $\lambda$.  However, near $\lambda=\lambda_1$ 
nearly degenerate perturbation theory is needed, giving
\begin{equation}
\epsilon_{1,2}(\lambda) \approx \frac{1}{2}\left[
(\sigma_1+\sigma_2)(\lambda-\lambda_1) \pm 
\sqrt{(\sigma_1-\sigma_2)^2(\lambda-\lambda_1)^2 + (\Delta
\epsilon)^2}\right].
\label{degen}
\end{equation}

It is helpful to consider {\em complex} values of
$\lambda$. Then the time evolution is no longer
unitary, and quasienergies are complex. They may be considered
as the $N$ values of a single $N$-valued analytic function, 
$\epsilon(\lambda)$,
with $N$ Riemann sheets connected at complex branch points\cite{kohn}. 
As is the case for real $\lambda$, the function $\epsilon(\lambda)$
is defined only modulo $\omega$, and we choose always
$0 \leq {\rm Re}\ \epsilon(\lambda)<\omega$. From (\ref{degen}) we
see that wac's for real $\lambda$
are manifested as branch points near the real axis of  
$\epsilon(\lambda)$, at
\begin{equation}
\lambda=\lambda_1 \pm \frac{\Delta\epsilon}{|\sigma_1-\sigma_2|} i .
\label{bpoint}
\end{equation}
(see Fig.~2).  When $\lambda$ passes along the real 
$\lambda$ axis through $\lambda_1$, 
as discussed above, there is a rapid change in the spatial part
of the two Floquet states, which are approximately interchanged.
In contrast, on a path 
starting from the real axis, going in a loop in the complex plane 
around the branch point (\ref{bpoint})
and back to the real axis, each of the two Floquet states returns 
approximately to its original spatial dependence.
Perturbation expansions in $\lambda$ 
have finite radii of convergence.  At $\lambda =0$ the
eigenstate $\varphi_j$ with energy $E_k$ defines
the quasienergy $\epsilon_j = E_k\pmod\omega $ 
on the j-th Riemann sheet of $\epsilon(\lambda)$ (the
indices $k$ and $j$ are unequal, in general, because $j$
labels increasing values of $\epsilon$ in $[0,\omega)$, whereas $k$
labels increasing energy values $E$ over $[0,\infty)$ ). 
Then the radius of convergence $\lambda_{c,j}$ for $\varphi_j$
and $\epsilon_j$ is the magnitude
of $\lambda$ at the branch point nearest the origin on that 
j-th sheet.

In the laboratory the perturbation, $\lambda H_1$, is commonly
turned on slowly. This can be characterized, as usual, by a 
switching  factor $e^{st}$ in the interaction ($s>0$):
\begin{equation}
H(t) = H_0 + \lambda e^{st} H_1(t)  \;\; {\rm where} \;\; H_1(t+T)=H_1(t) ,
\label{Hturnon}
\end{equation}
with the initial condition
\begin{equation}
\psi_j^{s,\lambda}(-\infty)=\varphi_j .
\label{asympsi}
\end{equation}
We are interested in the wave function 
$\psi_j^{s,\lambda}(t)$ at a specified time, say $t=0$.
This system is no longer periodic in time, and the solutions of the
time dependent Schr\"odinger equation corresponding to
(\ref{Hturnon}) and (\ref{asympsi}) are 
therefore no longer Floquet functions.
However, if the turn-on rate $s$ is slow compared to the
driving frequency $\omega$, it is useful to describe the solutions
at time $t$
in the basis of the Floquet functions (\ref{Floquet})
at the corresponding value of $\lambda(t)=\lambda e^{st}$. 
Two results are known\cite{holthaus}:

i) Every ac can be characterized by a rate
\begin{equation}
\xi=\frac{(\Delta\epsilon)^2}{ |\sigma_1-\sigma_2| \lambda_1} .
\label{xi}
\end{equation}
For large turn-on parameters $s \gg \xi$ (provided that $s \ll \omega$)
the solution will follow closely the initial Floquet state, as if there
were no ac. 
This is the so-called \cite{LanLif} Landau-Zener transition
through the ac.
On the other hand, in the adiabatic limit for this ac, $s \ll \xi$,
it will closely follow the Floquet state which passes the ac
continuously.
For $s \approx \xi$  the solution, after $\lambda(t)$ has passed
$\lambda_1$, will be a superposition of the two
Floquet states involved in the ac.

ii) The adiabatic theorem: 
Apart from an overall phase factor, the final state 
$\psi_j^{s,\lambda}(0)$ converges in the limit
$s \rightarrow 0$, and the limit is the Floquet state
of the periodic system that
corresponds to the quasienergy found by following the
quasienergy $\epsilon_j$ of the initial state 
$\varphi_j$ as a {\em continuous} function of $\lambda$.

These properties are well
established for Floquet systems with a finite basis. In the next
section we show that the limit $N\rightarrow\infty$ is
highly pathological.

\section{The Limit $N \rightarrow \infty$}

The general difficulties
raised by an infinite basis are outlined here; a detailed discussion
is given in the following two sections.

For definiteness we will
restrict the discussion to finite one-dimensional systems
with $x \in [-a,a]$ and
\begin{equation}
H(t) = p^2/2m + U(x) + \lambda V(x) \cos(\omega t), 
\label{Hmodel}
\end{equation}
where $U(x)$ and $V(x)$ are analytic and bounded.  Re~$\lambda$ will
be kept uniformly bounded, Re~$\lambda\leq\bar\lambda_1$, as $N$ 
becomes infinite.  For $\lambda=0$ 
the conventional energy spectrum $E_k$ is discrete, and for 
sufficiently high eigenvalues exhibits increasing spacings 
\cite{brenner} between
successive levels (approximately proportional to  
the level index, as for the case $U(x)\equiv 0$).  We will make 
repeated essential use of this feature.

In the limit
$N \rightarrow \infty$ one expects that for every $\lambda$ 
the $N$ quasienergies within the finite 
interval $[0,\omega)$ will form a dense spectrum.
In fact, Weyl showed\cite{weyl} that for $U(x)=\lambda=0$, so that
$E_k=(1/2m)(\pi/2a)^2k^2$, the spectrum is 
uniformly dense, provided only that $\omega/E_1$ is 
irrational.  This is the generic situation\cite{furst} 
for $U(x)$, $\lambda
\neq 0$.  Moreover, Howland \cite{howland} has shown that, for 
Hamiltonians of the form (\ref{Hmodel}), and for most 
values of the coupling strength
$\lambda$, the quasienergies have a dense point spectrum. 

What happens to the ac's in the limit $N \rightarrow \infty$? 
We find a very simple picture in the complex $\lambda$ plane:
ac's correspond to branch points, as in Fig.~2. 
With increasing
$N$ we add higher lying states of the unperturbed Hamiltonian to the 
basis and find that the gaps of newly introduced ac's, as well as 
the imaginary part of the corresponding branch points (see the Appendix),
decrease faster than any power law with $N$. Therefore the real 
$\lambda$ axis is a line of accumulation for the branch points.
Moreover, the branch points with Im~$\lambda > \bar\lambda_2$
are finite in number and tend to well-defined limits as 
$N \rightarrow \infty$, for any
$\bar\lambda_2>0$. The problems with the limit $N \rightarrow \infty$
are restricted to the immediate neighborhood of the real $\lambda$ axis.

For real $\lambda$  one expects the ac's to be dense in the
$\lambda - \epsilon$ plane. 
In fact, along each quasienergy curve $\epsilon(\lambda)$ obtained
within a finite basis approximation one expects a
dense set of ac's when $N\rightarrow\infty$. This has a 
number of consequences:

i) As there will be ac's on each quasienergy curve
for arbitrarily small values of
$\lambda$, the radii of convergence $\lambda_{c,j}$
of the perturbation theory in $\lambda$ for quasienergies 
and Floquet states all shrink to zero.

ii) The increase of the basis size from an initial value
$N$ will introduce
ac's arbitrarily close to any point on a given quasienergy 
curve defined with the initial basis of size $N$. Each of these ac's 
will make large changes in the corresponding Floquet states
over a finite $\lambda$-interval, strongly admixing and interchanging
pairs of states, as described above, 
and it is therefore
by no means clear whether or not Floquet states converge as 
$N \rightarrow \infty$.

iii) For any {\em finite} $N$ one can label the Floquet states 
$\varphi_n^{N, \lambda}$ and quasienergies $\epsilon_n^{N, \lambda}$
in such a way that they are continuous functions of $\lambda$
(if, as usual, we identify $\epsilon$ with $\epsilon+\omega$).
If we increase $N$, we have to rearrange labels for every
new ac, and it is not clear if there exists a meaningful labelling 
which tends to a well defined limit as $N \rightarrow \infty$.

iv) Let the periodic field be switched on over the time interval
$ -\infty < t \leq 0$,
with $\lambda(t) \equiv e^{st}\lambda $. For any {\em finite} 
$N$ there is a well defined adiabatic limit, as $s \rightarrow 0$, 
of the state (up to an overall phase factor) and of the quasienergy 
at the final time, $t=0$. 
Let $\xi_{\rm min}$ be the smallest of the 
rate parameters $\xi$, defined by (\ref{xi}), characterizing 
the relevant ac's in $[0,\lambda]$. Then, 
for $s \ll \xi_{\rm min}$ the solution of the Schr\"odinger 
Equation~(\ref{Hturnon}) will simply follow the Floquet state corresponding
to a continuous quasienergy curve. In the limit $N \rightarrow \infty$,
however, there will be ac's with arbitrarily small parameters $\xi$.
Therefore, for smaller $s$ more and more of these weak ac's will be
passed adiabatically, rather than undergo Landau-Zener transitions,
leading to completely different
final states $\psi_k^{s,\lambda}(t)$. Thus an adiabatic limit as 
$s \rightarrow 0$ cannot be expected.

\section{Convergence and Non-convergence of Floquet States}

For systems of type (\ref{Hmodel}) we shall present
strong arguments that, for a set of full measure
in $\lambda$ in an $N$-independent interval 
$0\leq\lambda\leq\bar\lambda$, the 
Floquet states converge in the limit
$N \rightarrow \infty$, even though the ac's are dense in the 
$\lambda-\epsilon$ plane.  This does not imply, however, that
the limit is a continuous function of $\lambda$.
For we will also argue that the 
ac's give rise to infinitely many $\lambda$'s, of measure zero
but dense in
any $\lambda$-interval, for which {\em none} of the Floquet
states of a finite basis converge. 
This can be stated more precisely as follows:

{\bf Proposition I (Convergence):}
For any interval $[\lambda_a,\lambda_b]$, any $\delta >0$, and
any $\eta >0$, there is an integer $M(\delta,\eta,\lambda_a,\lambda_b)$
with the following properties:
For any $M'>M$ and each $m=1,2,\dots,M$ there exists a label 
$m' \in \{1,2,\dots,M'\}$
and an overall phase factor $e^{i\alpha}$ such that
\begin{equation}
|\varphi_m^{M,\lambda}(x,t)
- e^{i\alpha} \varphi_{m'}^{M',\lambda}(x,t) | < \delta
\label{propIa}
\end{equation}
and
\begin{equation}
|\exp[-i(\epsilon_m^{M,\lambda}-\epsilon_{m'}^{M',\lambda})T]-1|
<2\delta ,
\label{propIb}
\end{equation}
for all $x$, for all $t \in [0,T]$, and
for all $\lambda \in [\lambda_a,\lambda_b]$, except for a subset 
of $\lambda$ of measure $\leq  \eta$.

{\bf Proposition II (Non-convergence):}
For any interval $[\lambda_a,\lambda_b]$, any $\delta >0$,
any $M$, any label $m \in \{1,2,\dots,M\}$,
sufficiently large $M' > M$ and any 
label $m' \in \{1,2,\dots,M'\}$,
\begin{equation}
\frac{1}{2a} \left|\int_{-a}^{a} dx \;\; \varphi_{m}^{M,\lambda}(x,t)^*
\varphi_{m'}^{M',\lambda}(x,t)\right| < \delta ,
\end{equation}
for a set of $\lambda$'s dense in $[\lambda_a,\lambda_b]$
and all $t$. That is, for this dense set of $\lambda$'s the 
eigenfunctions within the smaller basis 
have arbitrarily small overlap with any of those corresponding 
to the larger basis; they do not converge to a limit.

Proposition~I states that as $N$ grows, there is an 
increasing measure of $\lambda$ 
where a finite basis calculation gives the Floquet states
and quasienergies correctly within an arbitrarily small error.
For many practical purposes this supports the use of a finite basis
for describing a Floquet system. 
Nevertheless, it is important to realize,
as stated in proposition~II, that even for an
arbitrarily large basis size $N$ there are infinitely many 
$\lambda$'s, dense in any interval (albeit of total
measure zero), where a given
Floquet state $\varphi_{n}^{N}(t)$ does {\it not} converge
in the limit $N \rightarrow \infty$.

We can prove proposition~I for any system with the following model 
property which, we submit, 
captures for this purpose the essence of a real Floquet system of 
type (\ref{Hmodel}): For an ac at 
$\lambda_n \in [\lambda_a,\lambda_b]$ we again
define the interval $\lambda_n \pm w_n(\delta_n)/2$, outside of which
the admixture of the two unperturbed Floquet states changes 
by no more than $\delta_n$. Then 
we assume that we can choose a set $\{\delta_n\}$, with 
$\sum_{n=1}^{\infty} \delta_n < \delta$ for any chosen
$\delta>0$, such that 
$\sum_{n=1}^{\infty} w_n(\delta_n)$ converges.
This assumption appears to be satisfied in systems of type (\ref{Hmodel}), 
although we have no mathematical proof.  Numerical
calculations and analytical considerations (see Appendix)
suggest strongly that the gaps $\Delta\epsilon_n$ 
decrease faster than any negative
power of $n$, due to the increasing spacings of the unperturbed
energies $E_j$.  Let us choose $\delta_n$ to decrease relatively
slowly, as a small power of $n$, say
$\delta_n = \delta/(2n^2)$.  Then the sum over $w_n$ will converge
(see Eq. (\ref{width})), as long as the difference $|\sigma_1 - 
\sigma_2|$ in quasienergy slopes doesn't decrease too rapidly 
with $n$.  Numerical experience suggests this to be the case.

We make an argument based on the
Borel-Cantelli lemma~\cite{borel}.
Let us start, as usual, with an approximation to the system 
given by restriction
to a finite number $M$ of spatial basis functions.  The new
ac's introduced as this is increased to
a complete, infinite, basis set are labelled from $n_M$
to $\infty$.  For a large enough choice of the initial
basis size $M$, the partial sum
$\sum_{n=n_M}^{\infty} w_n(\delta_n)$,
which gives the measure of $\lambda$'s where a Floquet state
might be affected by more
than $\delta$ when increasing the basis size from $M$ to infinity,
can be made smaller than any given $\eta$.
For all other values of $\lambda$, those that are within none of the
intervals $\lambda_n \pm w_n(\delta_n)/2$, and which therefore
constitute a set of measure at least $\lambda_b-\lambda_a-\eta$, 
the Floquet states
are changed by no more than
$\sum_{n=1}^{\infty} \delta_n < \delta$.
This explains Eq.~(\ref{propIa}) of proposition I.

Equation~(\ref{propIb}) follows at once from Eq.~(\ref{propIa}).
The quasienergies can be determined from
$e^{i\epsilon_n^{M} T} = \varphi_n^{M}(x,t+T)/
\varphi_n^{M}(x,t)$.
For  Floquet states normalized by 
 $1/(2a) \int_{-a}^a dx |\varphi_n^{M}(x,t)|^2=1$,
we can choose an $x$ and $t$ where
$|\varphi_n^{M}(x,t)| \geq 1$. 
Then $\epsilon_n^{M'}$ can be determined from
$\varphi_n^{M'}(x,t)$, which differs from $\varphi_n^{M}(x,t)$
by less than $\delta$ (Eq.~(\ref{propIa})), leading to 
Eq.~(\ref{propIb}).

Proposition~II follows from the plausible, but unproved, assumption that
any quasienergy line of a finite basis ($M$) calculation will
show an ac within any given $\lambda$-interval, if the basis size
is increased sufficiently. 
(Failure of this assumption would imply that there do exist finite
$\lambda$-intervals in which a quasienergy line is {\em never}
crossed as $M \rightarrow \infty$.)
Using that assumption, we can argue straightforwardly:
Within any
given interval $I=[\lambda',\lambda''] \subset [\lambda_a,\lambda_b]$,
for a suitably large basis size $M$ one will eventually find an 
ac. This crossing changes the $n_1$-th Floquet state $\varphi_{n_1}^{M}$
by more than some chosen amount, say 40\% admixture of orthogonal
basis states, over some finite 
$\lambda$-interval.
Within that interval the two Floquet states of the ac then have
an overlap of less than $0.6$ with
$\varphi_{n_1}^{M}$.
We now select one of these states.
With a further increase in the basis
size, it will ultimately encounter an ac within the chosen
interval and will
be changed by more than 40\% over some smaller but still
finite $\lambda$-interval. 
Within that interval for an even larger basis size also the
second state 
will be changed by more than 40\% due to an ac.
This leads to four Floquet states within a finite
$\lambda$-interval, each with an overlap of less than $(0.6)^2$ with 
$\varphi_{n_1}^{M}$.
Repeating this argument sufficiently often,
one finds a basis size $M_1 > M$ and an interval $I_{n_1} \subset I$
where no Floquet state $\varphi_{n}^{M_1}$ has an overlap of more
than a given $\delta$ with $\varphi_{n_1}^{M}$ of the initial
basis size $M$, leading to proposition~II. 
As stated in proposition~I, these $\lambda$'s,
even though they are dense, have a measure which
tends to 0 as $M \rightarrow \infty$.

We can derive
a stronger version of proposition~II. 
Repeating the argument above for the $n_2$-th Floquet state
$\varphi_{n_2}^{M}$ of the initial basis size $M$, one finds
a basis size $M_2 > M_1 > M$ and an interval $I_{n_2}
\subset I_{n_1} \subset I$ where {\em no} Floquet state 
$\varphi_{n}^{M_2}$ has an overlap of more than a given $\delta$
with $\varphi_{n_2}^{M}$ nor with $\varphi_{n_1}^{M}$.
This argument can be repeated for all Floquet states
$\varphi_{n}^{M}$ of the initial basis.  Thus there is a dense
set of $\lambda$'s in any $\lambda$-interval,
where {\em none} of the Floquet states of a finite basis 
converges \cite{fnote}.

\section{ Functional Dependence on Perturbation Strength; Adiabatic Limit}

\subsection{Labelling of Floquet States}

Proposition~II of the last section has an immediate consequence:
A continuous labelling of the Floquet states $\varphi_n^{N,\lambda}$
and quasienergies $\epsilon_n^{N,\lambda}$
as a function of $\lambda$, possible for finite $N$, is
no longer possible in the limit $N \rightarrow \infty$. 
We therefore propose a new way of labelling, 
useful at least for small $\lambda$:
We assign the label $n$ to a Floquet state,
if its overlap with the $n$-th eigenstate
$\varphi_n(x,t)$ of $H_0$ is larger than 50\%, in the sense
\begin{equation}
\frac{1}{T} \int_0^T dt \;\;
\frac{1}{2a} \int_{-a}^a dx \;\; |\varphi_n(x,t)^*
\varphi_n^{\lambda}(x,t)| \;\; > \;\; 0.5   .
\end{equation}
This procedure will not always find a label for a Floquet state.
This is obvious for large $\lambda$, where
none of the Floquet states resembles a low lying unperturbed state.
But, even for small $\lambda$, in the center region of an 
ac the overlap with an unperturbed state will be less than 50\%.

Since for small $\lambda$ the quasienergies 
are very flat as a function of $\lambda$, and ac's
will generically occur only between Floquet states related to
states of $H_0$ that are far apart in energy, their 
ac's are expected to have a width that decreases faster than 
any power law
as $\lambda$ goes to zero (Appendix). We therefore expect that,
on the interval $[0,\bar\lambda]$, the labelling works 
for a Cantor set of $\lambda$ values with 
finite measure less than $\bar\lambda$, and that this measure 
approaches $\bar\lambda$
as $\bar\lambda \rightarrow 0$.

\subsection{Adiabatic Turn On}

Even though perturbation theory
for Floquet states and quasienergies in $\lambda$ has zero
radius of convergence in the limit $N \rightarrow \infty$, 
we find a simple, strict result
for the perturbation expansion of a solution $\psi_k^{s,\lambda}(t)$
of the Schr\"odinger equation~(\ref{Hturnon}) , where the periodic driving
is turned on from $t=-\infty$ by the factor $\lambda e^{st}$ (we 
remark again that 
$\psi_k^{s,\lambda}(t)$ is {\it not} a Floquet state).

{\bf Theorem:}
The perturbation expansion of $\psi_k^{s,\lambda}(t)$
in $\lambda$, for $t<\infty$, has an infinite radius
of convergence for any $s>0$, i.e.\ $\psi_k^{s,\lambda}(t)$ is
an entire function of the complex variable $\lambda$. 

 This can be proven by majorizing the perturbation expansion. 
With the wave function $\chi(t)$ expressed in the
interaction picture,
\begin{equation}
\chi(t)=e^{iH_0 t} \psi_k^{s,\lambda}(t),
\end{equation}
the Schr\"odinger equation (\ref{SE}) becomes
$i d\chi(t)/dt = \lambda W(t) \chi(t)$, with
$W(t)=e^{iH_0 t} e^{st} V(x) \cos(\omega t) e^{-iH_0 t}$.
The familiar formally iterated solution is
\begin{eqnarray}
\chi(t) = \biggl[ 1 & + & (-i\lambda) 
\int_{-\infty}^t dt_1 W(t_1)
+ (-i\lambda)^2 
\int_{-\infty}^t dt_2 \int_{-\infty}^{t_2} dt_1 W(t_2) W(t_1)
\nonumber \\
& + &  (-i\lambda)^3 
\int_{-\infty}^t dt_3 \int_{-\infty}^{t_3} dt_2
\int_{-\infty}^{t_2} dt_1 W(t_3) W(t_2) W(t_1) + \dots \biggr]
\;\; \chi(t=-\infty)
\label{chipert}
\end{eqnarray}
Since $e^{-iH_0 t}$ is unitary,
it can easily be shown that for any normalized states $f$ and $g$
\begin{equation}
<f|W(t_n) W(t_{n-1}) \dots W(t_2) W(t_1) |g> \;\;\; \leq \;\;
V_{\rm max}^n e^{s(t_n+\dots+t_1)} ,
\end{equation}
where $V_{\rm max}$ is the maximum of $|V(x)|$. From this
the $n$-th order term of the perturbation expansion of
$<f|\chi(t)>$ can be majorized by $(1/n!) (\lambda V_{\rm max}/s)^n
e^{st}$ and thus the expansion~(\ref{chipert}) converges for any
$s>0$.

That is, the state which evolves with a given switching-on rate $s$  
is uniquely and well defined (in a finite {\em or}
infinite basis), for an arbitrarily large final interaction strength,
in spite of the convergence problems with Floquet states. There 
{\em is}, however, an anomaly: there is no well-defined adiabatic 
limit when $s\rightarrow 0$ (see also Sec. III).
 
Although an adiabatic limit in the usual sense does not exist,
we now show that for sufficiently small $\lambda$ 
there is a (logarithmically) large window of turn-on parameters 
$\underbar{s} < s < \bar{s}$,
where the final state $\psi_k^{s,\lambda}(t)$ is {\em almost}
independent of $s$.

{\bf Proposition III:}
For any small $\delta >0$ and large $\eta >0$ there 
exists a $\bar\lambda >0$
and an interval $[\underbar{s},\bar{s}]$ with
\begin{equation}
\bar{s}/ \underbar s > \eta ,
\label{ratio}
\end{equation}
such that for all $s',s'' \in [\underbar{s},\bar{s}]$, all
$\lambda < \bar\lambda$, all $x$ and all states $k$,
\begin{equation}
\frac{|\psi_k^{s',\lambda}(x,t=0) - e^{i\alpha(s',s'')}
\psi_k^{s'',\lambda}(x,t=0)|}{\lambda}
< \delta ,
\end{equation}
where $e^{i\alpha}$ is a (physically uninteresting)
overall phase factor.

That is, for any desired level of convergence (as defined by $\delta$)
we can find a range of $s$ of arbitraily large relative size (arbitrarily
large $\bar s/\underbar{s} = \eta$) by restricting $\lambda$ to a
sufficiently small value.

Consider any finite $N$ and the Floquet state arising out of
the unperturbed state $\varphi_k$. The perturbation series~(\ref{chipert})
involves first order terms of the form
\begin{equation}
\chi^{s,(1)} \equiv \lambda \sum_{k'} \frac{V_{k',k}}
{E_{k'}-E_k \pm \omega -is} \varphi_{k'} .
\label{chifirst}
\end{equation}
 From this it is clear that to achieve independence of $s$ with
an accuracy $\delta$,
$s$ must be smaller than a value $\bar s$ given by
\begin{equation}
\bar{s} \lambda \sum_{k'} \biggl| \frac{V_{k',k}}
{(E_{k'}-E_k \pm \omega)^2} \varphi_{k'} \biggr| \;\; < \;\; \delta .
\label{supper}
\end{equation}
For small enough $\lambda$ all higher order terms up to any
finite order $n$ may be neglected (assuming, as we do, that there
is no exact resonance $E_{k'}-E_k = \pm m\omega$, $m \leq n$).

However, for any given $\lambda_0$, no matter how small, there will exist
a sufficiently large basis size such that there are some (weak)
ac's associated with {\em any} quasienergy     
in the $\lambda$-interval $[0,\lambda_0]$.  Consider for any initial
basis size $N_0$ an arbitrary quasienergy curve $\epsilon_k(\lambda)$,
continuous over the interval $0\leq\lambda\leq\lambda_0$.  As the
basis size is then increased to a sufficiently large size $N'>N_0$,
avoided crossings of $\epsilon_k(\lambda)$ will be introduced  within
this interval.  These ac's will be characterized by rates 
(see Eq.~(\ref{xi})) which we label 
$\xi_1, \xi_2, \dots$ , in order of increasing energy of the states
at $\lambda =0$ from which the crossing curves arise,
which then assures that they are ordered with decreasing rates: 
$\xi_{i+1} < \xi_i$.
As $N \rightarrow \infty$ and $\lambda(t)$ ($\equiv
\lambda e^{st}$) grows from 0 to $\lambda$ all these
(infinitely many) weak ac's will be encountered,
and each of them gives rise to an admixture of a new state into
$\psi_k^{s,\lambda}(x,t=0)$ with amplitude (see~\cite{LanLif})
smaller than
\begin{equation}
\delta_n = \sqrt{\pi\xi_n/s}. 
\end{equation}
To make the total variation in $\psi_k^{s,\lambda}(x,t=0)$, 
as defined by $\sum_n \delta_n$, 
smaller than a given $\delta$ over the whole range of turn-on 
rates set by Eq.~(\ref{ratio}), we use the
following properties of the $\xi$'s in the limit $\lambda
\rightarrow 0$:
\begin{equation}
\lim_{\lambda \rightarrow 0} \xi_1 = 0
\label{xi1}
\end{equation}
\begin{equation}
\lim_{\lambda \rightarrow 0} \xi_{l+1}/\xi_l = 0 
\label{xiratio}
\end{equation}
Equation~(\ref{xi1}) follows from the fact, that for small enough
$\lambda$ the largest ac of $\epsilon_k^{\lambda}$ in 
$[0,\lambda]$ will be with an arbitrarily high-lying state
$k'$ of $H_0$ and that the rate $\xi_1$
of the ac, according to Eq.~(\ref{xi}) (see also the Appendix), decreases 
faster with $k'$ than any power law.
Equation~(\ref{xiratio}) is due to the fact, that the ratio
$\xi_{l+1}/\xi_l$ is, in the most unfavorable case, due to ac's of 
neighbouring levels $k'+1$ and $k'$
of $H_0$ with $\epsilon_k^{\lambda}$. 
The ratio of the corresponding
quasienergy splittings decreases exponentially with $(k'+1)^2-k'^2
= 2k'+1$ and goes to zero in the limit $\lambda \rightarrow 0$.

 From these properties we conclude that for $s$ larger than any
$\underbar{s} <\bar{s}/\eta$ the variation in 
$\psi_k^{s,\lambda}(x,t=0)$ due to infinitely many ac's
can be made smaller than any given $\delta$ for 
sufficiently small $\bar\lambda$.
Therefore for small enough $\bar\lambda$ one finds 
for all $\lambda \leq \bar\lambda$ almost adiabatic behaviour
in a window $[\underbar{s}, \bar{s}]$ of turn-on parameters
with $\bar{s}/\underbar{s} $ arbitrarily large.

\subsection{Conservation of Quasienergy}

Here we consider again the limit of turning on $\lambda(t)$ arbitrarily
slowly from $0$ up to some arbitrary $\lambda$.
We have seen
that the eigen{\em function} $\psi_k^{s,\lambda}(0)$ does not
have a limit for $s \rightarrow 0$.  However, 
we will argue here (but not prove mathematically) that,
as $s \rightarrow 0$,
$\psi_k^{s,\lambda}(0)$ is within arbitrary accuracy
a linear combination of Floquet states of the periodic
Hamiltonian (Eq.~(\ref{Hperiodic})) which have 
quasi{\em energies} arbitrarily
close to the {\em initial} value $\epsilon_k(\lambda=0)$ 
--- {\it i.e.}, the
{\it energy} modulo $\omega$. Thus in the
limit $s \rightarrow 0$ quasienergy is a conserved quantity:

{\bf Proposition IV:}
For any $\delta >0$ and any $\lambda >0$ there exists an $s_0>0$
such that for all $s<s_0$
\begin{equation}
\int_{-a}^{a} dx |\psi_k^{s,\lambda}(x,t) - e^{i\epsilon_k T} 
\psi_k^{s,\lambda}(x,t+T)|^2 < \delta
\end{equation}
for all $t \leq 0$, and for all states $k$. 
This is true even though $\lim_{s \rightarrow 0} \psi_k^{s,\lambda}(x,t)$
does not exist for any finite time $-\infty < t \leq 0$. 

To make this plausible we will use a simplified geometrical picture,
in which quasienergies are linear functions of $\lambda$. Although all
ac's are now represented by actual crossings, they will be traversed
dynamically like real ac's with some finite rate parameters $\xi$.
The infinitely many quasienergy lines fall into
two classes, according to whether the magnitudes of their slopes
are smaller or larger than a specified critical value 
$\sigma_c=\delta'/(2\lambda)$.
We will further assume in this model that the latter (large slope) class,
for any $\delta'>0$ and any $\lambda >0$, has only a finite number 
$K(\delta',\lambda)$ of members. 

In the course of turning on the periodic driving from 0 up 
to $\lambda$ with 
a small enough turn-on parameter $s$, the infinitely many quasienergy 
lines with slopes $\leq \sigma_c$ will, at most, cause a deviation
in quasienergy by $\sigma_c \lambda$ ($= \delta'/2$) away from the initial
value $\epsilon_k(\lambda=0)$. The additional changes in
quasienergy from ac's with the finite number $K(\delta',\lambda)$
of steeper quasienergy lines can also be restricted to be less
than $\delta'/2$ by a sufficiently small choice of turn-on rate
$s$:  Since the ac's are dense on any quasienergy line,
for sufficiently small $s$ one will be diverted from any of the 
steep quasienergy lines within any given
small quasienergy range, which we choose to be $\delta'/2K$.
The total change in quasienergy as $\lambda$ is increased from 0
to its final value thus can be made smaller than $\delta'/2
+ K\delta'/(2K) = \delta'$ for
any given $\delta'$.
For $\delta'=\sqrt{\delta}/T$ and
within this simplified model we have thus proved proposition~IV.

The same arguments should hold for the original Floquet problem, as
including the quasienergy dependence on $\lambda$ at the
ac's reduces their steepness and thus further
reduces the spread in quasienergy from the above estimates.
Also, the overall nonlinear dependence on $\lambda$ which follows, 
{\it e.g.}, from
second order perturbation theory, poses no problem for the argument.
The assumption that the number $K(\delta',\lambda)$ of slopes
larger in magnitude than $\sigma_c$ is finite, remains reasonable,
since an increase in the basis size introduces
Floquet states originating from higher lying states of $H_0$,
which show decreasing dependence on $\lambda$.
However, the problem of defining slopes at all
(in the limit $N \rightarrow \infty$), as there is no continuous
labelling of the quasienergies as a function of $\lambda$,
will make a mathematical proof difficult.

 This proposition leads us to a better understanding of the nature 
of the state $\psi_k^{s,\lambda}(0)$ in the limit
$s \rightarrow 0$. 
As $\psi_k^{s,\lambda}(0)$ is a linear combination of Floquet states
with quasienergies closer and closer to $\epsilon_k$,
it changes constantly as $\delta$ and $s$ go to zero. Thus, while 
there cannot be an adiabatic limit for the wave function, 
quasienergy is conserved in the limit $s\rightarrow 0$.

\section{The Status of Traditional Finite Order Perturbation Theory}

Non-linear optics is a major field of science in which traditional,
finite order perturbation theory in the applied electric field
(usually to low order) successfully describes experiment.
Here we shall show why this well-established theory is
consistent with our considerations in spite of our conclusion that,
strictly speaking the radii of convergence, $\lambda_{c,j}$, of
perturbation theory vanish.

We reiterate first that if the turn-on rate 
$s \rightarrow 0$, perturbative non-linear optics in fact fails.
For small $\lambda$ this failure is due to near-resonances,
$E_{k'}-E_k \approx \pm n\omega$, generally with very high-lying 
excited states.
(This is the reason why, for any finite basis size $N$, 
$\lambda_c$ is finite.)

When the perturbation is turned on as in Eq.~(\ref{Hturnon}),
we have seen that, provided the turn-on rate $s$ is small enough
but exceeds a lower limit, $\underbar{s} $, then as $\lambda
\rightarrow 0$, the resultant state can be made arbitrarily
close to the traditional first order perturbative solution.
A similar result can be derived for the traditional 
perturbative solution up to any finite order. In typical
laboratory situations we have seen that $\underbar{s} $
is exponentially and unphysically small.  For finite small
$\lambda$ the perturbation expansion is asymptotically
convergent.

Finally we briefly mention the unavoidable effects of line broadening.
The quantum system of interest is inevitably in contact with its
environment, and there are interactions between the
many particles that ordinarily constitute the quantum system of
interest, so the individual particle states are lifetime broadened.
We conjecture that if broadening is characterized by a finite width 
$\Gamma$, then a finite radius of convergence will be
restored.

\section{Conclusion}

There have been two standard approaches to dealing with the
behavior of quantum systems subject to strong time periodic
fields.  One is the use of finite order perturbation theory
(e.g., second or third order nonlinear optical susceptibilities),
and the other the exact solution of the problem within a
finite basis of states.  But both of these approaches miss
qualitative features of the exact mathematical solutions.

We have shown by a set of ``propositions" (as opposed to rigorous
mathematical proofs) that for a large class of time periodic problems
the structure of the exact states and the
quasienergy spectrum is remarkably irregular. By ``exact" we mean
here that the complete infinite set of basis states is included. 
Interaction with the environment is neglected.  
We have considered the states and 
quasienergies as functions of the strength $\lambda$ of the 
time periodic potential, as the number $N$ of basis states
becomes infinite.  We have found that
in any interval $\lambda_a < \lambda < \lambda_b$, although the
states converge to a well defined limit as $N\rightarrow\infty$
for a set of $\lambda$ with the full measure
$\lambda_b-\lambda_a$ of the interval, there is a  set of
$\lambda$, of total measure zero, but {\em dense} 
within every finite interval, for which the states do {\em not} converge.
As a result, in contrast to the situation for any finite $N$,
it is impossible to label states and quasienergies continuously
as a function of $\lambda$.  The familiar quasienergy ``dispersion"
curves as functions of $\lambda$ (as shown, {\it e.g.}, in Fig. 1) become
 discontinuous everywhere.  One consequence of these
discontinuities is the absence of a true adiabatic limit; there
is no unique final state to which the system tends as the periodic
perturbation is switched on arbitrarily slowly.

But these pathologies, including a radius of convergence of
perturbation theory in $\lambda$ which approaches zero as 
$N \rightarrow \infty$, do not show up under most physically 
realistic circumstances.  In particular,   
we have explained
the familiar and well established success of ordinary time dependent
perturbation theory in terms of the modified adiabatic theorem and
the typical smallness of the parameter $ \underbar{s} $ which enters
that theorem in practice, as well as the successes of finite basis 
calculations.

\acknowledgements

This work was supported by the NSF under Grant No. PHY94-07194 and
DMR96-30452, as well as the Deutsche Forschungsgemeinshaft. We 
profited from discussions 
with S. Fishman, H. Metiu, and F. Pikus.  One of us (WK) thanks
Prof. J. Moser for a helpful conversation.  The work was
stimulated in part by the experiments of M. Sherwin on the
response of electrons in semiconductor quantum structures to
intense far infrared laser fields.

\newpage

\appendix
\section*{Exponential Decrease with Basis Size N of Newly Introduced 
Quasienergy Gaps}

We label the eigenstates of the time independent Hamiltonian 
($\lambda=0$) by an index $j$ which increases with the unperturbed 
energy.  We consider 
the solution of the full time dependent problem in the limited spatial 
basis of the first N such states.  We demonstrate here that, for 
sufficiently small $\lambda$, the new ac's introduced 
by the inclusion of the next basis state (labelled $N+1$) are 
characterized by quasienergy gaps that are smaller than a bound 
which decreases exponentially with $N$.

As in Sec. II, let us take as the ``$N^{th}$ level" Hamiltonian 
the representation of $H(t)$ in the basis of the first 
$N$ unperturbed states:
\begin{equation}
H^N(t) \equiv H_0 + \sum_{j,\ell=1}^N|\ell\rangle\langle
\ell|V(t)|j\rangle\langle j| \equiv
H_0 + V^N(t) ,
\label{HN}
\end{equation}
where the unperturbed Hamiltonian is
\begin{equation}
H_0 = \sum_{j=1}^\infty E_j^0 |j\rangle\langle j|.
\label{h0}
\end{equation}
Since $H^N$ has the same time periodicity as the full Hamiltonian, 
the corresponding time dependent Schr\"odinger equation has solutions 
of the standard Floquet form (\ref{Floquet}),
\begin{equation}
\psi_k^N(x,t) = \exp(-i\epsilon_k^N t)u_k^N(x,t),
\label{psik}
\end{equation}
with $k=1,2,\ldots N$, 
where the functions $u_k^N(x,t)$ are time periodic.  Therefore, the 
solutions formed from the basis of the first $N$ states are of the form
\begin{equation}
|u_k^N\rangle = \sum_{j=1}^N\sum_{n=-\infty}^{\infty}a_k^N(j;n)
e^{-in\omega t} |j\rangle ,
\label{defa}
\end{equation}
where normalization imposes the restriction
\begin{equation}
\sum_{j=1}^N\sum_{n=-\infty}^{\infty} |a_k^N(j;n)|^2 = 1 .
\label{norm}
\end{equation}
Now we include in (\ref{HN}) the next highest state 
$|N+1\rangle$, whose unperturbed 
energy $E_{N+1}^0$ can be written as
\begin{equation}
E_{N+1}^0 = \epsilon + M_N\omega ,
\end{equation}
with $\epsilon$ confined to the fundamental stripe, 
$0\leq\epsilon<\omega$. At some value of the coupling $\lambda$ 
the quasienergy of this state, $\epsilon$, may equal that of   
one of the solutions labelled k ($1\le k\le N$) 
in the basis of the first $N$ states (see Eqs. (\ref{psik}) and 
(\ref{defa})).  The 
perturbation of the remaining potential, $V-V^N$, turns that into an 
{\em avoided} crossing, with a gap given approximately by 
twice the corresponding matrix element,
\begin{equation}
\Delta = \lambda\sum_{j=1}^N\langle N+1|V|j\rangle\left[ a_k^N(j;M_N+1) 
+ a_k^N(j;M_N-1) \right].
\label{gap}
\end{equation}
We now place strong limits on the size of the right hand side of 
this equation. Since the state label $k$ and the basis size $N$ 
will remain fixed, for simplicity of notation we will no longer 
write the subscript $k$ and superscript $N$ on the coefficients 
$a(j;n)$.  The time dependent Schr\"odinger equation for $u(x,t)$ 
can be rewritten as a set of equations for these coefficients:
\begin{equation}
[n\omega + \epsilon - E_j^0]a(j;n) = \frac{\lambda}{2}\sum_{\ell=1}
^N\langle j|V|\ell\rangle [a(\ell;n+1)+a(\ell;n-1)],
\label{eqna}
\end{equation}
for $j = 1, 2, \ldots N$.
We emphasize that this is the {\em exact} equation for the  
time dependent problem in the finite basis $N$.  It contains all 
orders of $\lambda$ and makes no reference to convergence of 
perturbation theory; there may be arbitrary resonances or near 
resonances of the time dependent Hamiltonian between the initial 
$N$ states.  We will draw only on the fact that normalized 
solutions, satisfying (\ref{eqna}), exist.  
From Eq. (\ref{gap}) we see that we need the 
coefficients $a(j;n)$ only for the large values of frequency index 
$n\approx M_N$.  The high lying energies of the unperturbed static 
Hamiltonian are approximately $E_j^0\approx j^2\omega_0$ (where
$\omega_0 = (1/2m)(\pi/2a)^2$).    Then 
$M_N\omega + \epsilon = E_{N+1}^0\approx (N+1)^2\omega_0$, and the 
factor in square brackets on the left hand side of Eq. (\ref{eqna}) 
for the case of interest, $n=M_N$, is greater than $2N\omega_0$ 
for any value of $j$ (the smallest value occurs for the largest 
possible index, namely $j=N$). We use the symbol $V_0$ to denote 
the maximum  absolute value of the matrix elements of $V$ between 
{\em any} two basis states.  Then the absolute value of the 
coefficient $a(j;n)$ is limited by Eq. (\ref{eqna}) to:
\begin{equation}
|a(j;M_N)| \leq \frac{\lambda V_0}{4N\omega_0}\sum_{i=1}^N[ |a(i;M_N+1)|
 + |a(i;M_N-1)| ] \leq \frac{\lambda V_0}{2\sqrt{2N}\omega_0},
\label{mnlim}
\end{equation}
where we have used only the limitation imposed by normalization, Eq. (\ref{norm}), 
on sums over the absolute values of any subset of the 
coefficients $a(j,n)$ corresponding to a single state $u_k^N$, namely 
$|a_1|+|a_2|+\ldots|a_m|\le \sqrt{m}$.  But we can do much better, 
essentially by iterating this process.  We start with Eq. (\ref{eqna}) 
for a smaller value of the photon index, $n=M_N-p$, use the argument 
just given to limit the right hand side for the next higher value, 
$n=M_n-p+1$, and work back to the value of interest, $n=M_N$. We 
choose the starting integer $p$ as the integer part of 
$N\omega_0/\omega$ (this gives a substantial improvement only for 
$N\omega_0/\omega\gg 1$, so we choose $N$ large enough for this 
to be the case). Then the coefficient in square brackets on the 
left side of (\ref{eqna}) is greater than $N\omega_0$.  Thus, by 
exactly the same kind of argument that led to (\ref{mnlim}) we have
\begin{equation}
|a(j;M_N-p)| \leq \frac{\lambda V_0}{\sqrt{2N}\omega_0}.
\label{plim}
\end{equation}
 The same limitation holds for $|a(j;M_N-p+2q)|$, with 
$q=1,2,\ldots,p$, where the
coefficient on the left of (\ref{eqna}) is even larger.  Then we use
these maximum values to bound the right hand side of (\ref{eqna})
for the next iteration, for the values of $n$ lying between those
just limited: $n=M_N-p+2q-1$ with $q=1,2,\ldots,p$ ({\em not} for
$q=0$):
\begin{eqnarray}
|a(j;M_N-p+q)| \leq &&\frac{\lambda V_0}{2N\omega_0}\sum_{i=1}^N
\left[ |a(i;M_N-p+q+1)| + |a(i;M_N-p+q-1)| \right] \nonumber\\
&&\leq \frac{1}{\sqrt{2N}}\left[\frac{\lambda V_0}{\omega_0} \right]^2,
\end{eqnarray}
where the first inequality comes again directly from (\ref{eqna}) 
with the minimum possible coefficient on the left hand side, and 
the final inequality results from substitution of (\ref{plim}), 
which holds for each value of the state index $i$, into the middle 
expression. This is repeated, using these bounds to limit the right 
hand side of (\ref{eqna})
for the $n$ values lying between {\em these}:
\begin{equation}
|a(j; M_N-p+2q)| \leq 
\frac{1}{\sqrt{2N}}\left[\frac{\lambda V_0}{\omega_0} \right]^3,
\end{equation}
now for $q=1,2,\ldots,p-1$.
We then repeat this $p-2$ times, with the power of $(\lambda V_0/  
\omega_0)$ increasing by one and the range of $q$ decreasing by
one at each iteration, to obtain
\begin{equation}
|a(j;M_N)| \leq \frac{1}{2N}\left[\frac{\lambda V_0}
{\omega_0} \right]^{p+1} \le \frac{1}{\sqrt{2N}}
\left[\frac{\lambda V_0}{\omega_0} \right]^{N\omega_0/\omega}.
\label{lima}
\end{equation}
Finally, we use this in Eq. (\ref{gap}) to put limits on the size 
of the gap:
\begin{equation}
\Delta \le \lambda V_0\sqrt{2N}\left[\frac{\lambda V_0}
{\omega_0}\right]^{N\omega_0/\omega}.
\end{equation}
Therefore, the gap is limited by this to be exponentially decreasing 
with basis size $N$, at least for small enough coupling, 
$\lambda < \omega_0/V_0$.

For specific examples of the spatial dependence of the time 
dependent potential $V(x)\cos\omega t$ we can construct even 
tighter limits.  There are, in particular, two limiting cases 
of interest: with the square well confining potential in the 
interval $-a<x<a$ we take (i) $V(x) = 2V_0\sin(\pi x/2a)$, or
(ii) $V(x) = V_0a\delta(x)$. In  both cases the only non-zero 
matrix elements of $V(x)$ between the eigenstates of the square 
well are of magnitude $V_0$.  In the first instance (sinusoidal 
potential) these occur only for nearest neighbor states in the 
energy ladder, $\langle j|V|k\rangle = V_0\delta_{k,j\pm 1}$, 
whereas for the delta function potential all pairs of even 
spatial parity states are connected by $V_0$ regardless of how 
far apart in energy they are (and all odd parity states with 
vanishing wave function at $x=0$, are, of course, totally unaffected 
by the potential).   These are then limiting cases of short and 
long range effects of the spatial potential relative to the 
energy spectrum of the unperturbed static square well potential. 

For the sinusoidal potential the right hand side of Eq. (\ref{eqna}) 
contains only the four coefficients corresponding to $l=j\pm 1$, 
so that the inequality (\ref{mnlim}) becomes
\begin{equation}
|a(j;M_N)| \le \frac{\lambda V_0}{2N\omega_0}.
\end{equation}
The limitation to four coefficients occurs at each stage of the 
iterative process which led to (\ref{lima}), which limit now becomes
\begin{equation}
 |a(j;M_N)|  \le \frac{1}{N}\left[\frac{\lambda V_0}
{N\omega_0} \right]^{N\omega_0/\omega}.
\end{equation}
Note that by choosing $N$ large enough ($\lambda V_0/N\omega_0<1$, 
as well as $N\omega_0/\omega>1$) we find ultimate exponential 
(indeed, powers of $(1/N!)$) decrease of the gaps for arbitrarily 
large coupling strength $\lambda$ for this case.

We also can obtain tighter limits for the delta function potential. 
In this case we can rewrite Eq. (\ref{eqna}) as
\begin{equation}
a(j;,n) = \frac{\lambda V_0/2}{n\omega+\epsilon-E_j^0} \left[A_N(n+1)
 + A_N(n-1) \right],
\label{adelt}
\end{equation}
where we have defined
\begin{equation}
A_N(n) = \sum_{i=1}^N a(i;n),
\label{defA}
\end{equation}
and throughout the analysis of this case the unperturbed eigenstate 
index $j$ refers only to even parity states (we have noted above that 
the odd parity states are not affected by this potential).
Then we can sum (\ref{adelt}) over the eigenstate index $j$ to find 
a recursive relationship for the $A_N(n)$:
\begin{equation}
A_N(n) = (\lambda V_0/2\omega_0) S_N(n) [A_N(n+1) + A_N(n-1)],
\label{eqA}
\end{equation}
where we have defined one more sum,
\begin{equation}
S_N(n) = \sum_{j=1}^N \left[ \frac{n\omega+\epsilon}{\omega_0} 
- j^2 \right]^{-1} \equiv \sum_{j=1}^N \frac{1}{C_n^2 - j^2} \approx 
\frac{1}{C_n} \ln \frac{C_n+N}{C_n-N}.
\label{sn}
\end{equation}
The final approximation on the right hand side is the Euler-Maclaurin 
integral estimate for the sum; corrections are of order $1/C_n$.  
Now, as before, we start by considering the recursion relation 
(\ref{eqA}) for $n=M_N - p$, with $p$ the integer part of 
$N\omega_0/\omega$ and limit the right hand side by the maximum 
imposed by the normalization condition: $|A_N(n)| < \sqrt{N}$, so that
\begin{equation}
|A_N(M_N-p)| < \frac{\lambda V_0}{2\omega_0}\frac{\ln(4N+1)}{\sqrt{2N}}.
\end{equation}
We use this in the right hand side of (\ref{eqA}) for the next higher 
value of $n$, namely $n=M_N-p+1$, and iterate as before to obtain
\begin{equation}
|A_N(M_N)| < \left[\frac{\lambda V_0\ln (4N+1)}{ 2\omega_0\sqrt{2N}}
\right]^{ N\omega_0/\omega}.
\end{equation}
Finally, we put this back into the equation (\ref{adelt}) for the 
original coefficient $a(j;M_N)$ to find
\begin{equation}
|a(j;M_N)| < \frac{\lambda V_0}{(2N+1)\omega_0}\left[
\frac{\lambda V_0\ln (4N+1)}{ 2\omega_0\sqrt{2N}}
\right]^{ N\omega_0/\omega}.
\end{equation}

\begin{figure}
\caption{
Quasienergy spectrum 
as a function of $\lambda$ for $N=10$ for the free
particle in a box with harmonic driving and frequency $\omega=8.3$
(see Eq.~(5)).
One finds many avoided crossings, with a typical one marked by the
dashed box.
There are some strictly {\em real} crossings, corresponding to states of
opposite parity under the combined symmetry operation of 
spatial inversion plus time translation by half a period
$T/2$.  Other apparently real crossings are just so weakly avoided
that they can't be resolved.
}
\end{figure}

\begin{figure}
\caption{
Position of branch points in the complex $\lambda$ plane for system and
parameters of Fig.~1.
Just one quadrant is shown, as the position of branch points is symmetric
with respect to the axes Re$(\lambda)=0$ and Im$(\lambda)=0$, since
$\epsilon(\lambda^*)=\epsilon^*(\lambda)$ and $\epsilon(-\lambda)=
\epsilon(\lambda)$.  There are no branch points on the real axis, but
they do appear on the imaginary axis, as shown.
}
\end{figure}


\begin{references}
\bibitem{fnote} 
This non-convergence implies\cite{fishman} 
that the corresponding quasienergy spectrum is not a point 
spectrum for these values of $\lambda$.  But Howland has 
shown\cite{howland} that the spectrum is also
not absolutely continuous. It follows that the quasienergy
spectrum is singular continuous for a set of $\lambda$ which is  
everywhere dense but of total measure zero.  See also the conclusion
of G. Casati and I. Guarneri, Comm. Math. Phys. {\bf 95}, 121
(1984), that the quantum kicked rotor has singular continuous quasienergy spectra for a non-empty set of driving frequencies.
\end{references}
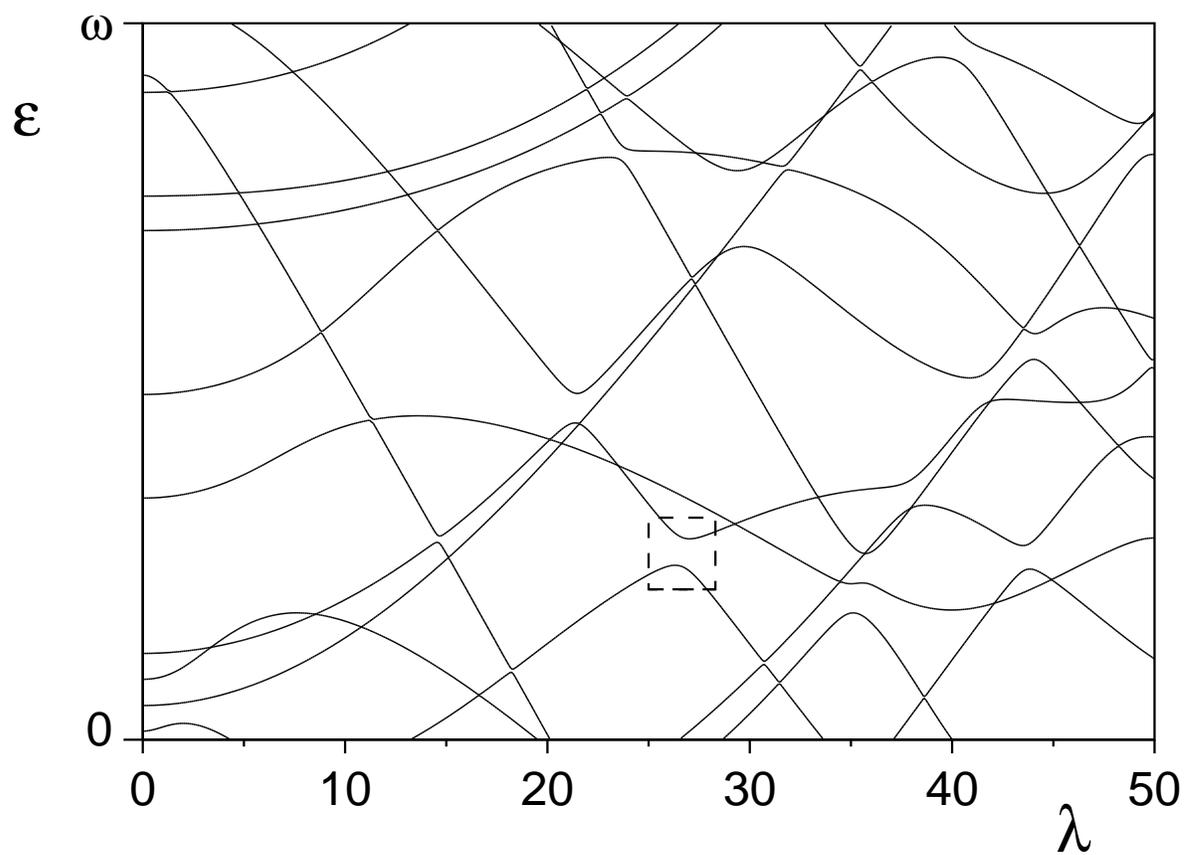
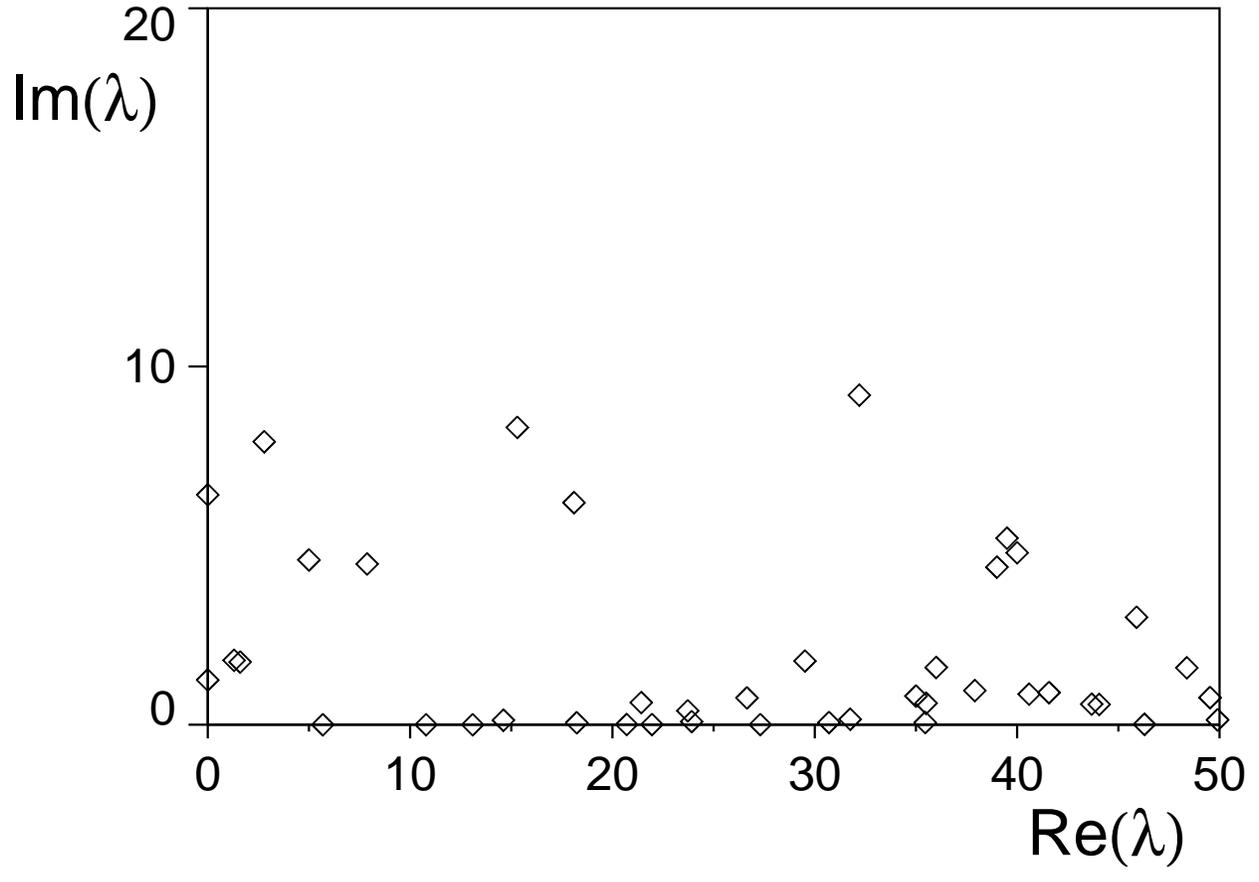
\end{document}